\documentclass[11pt]{article}
\usepackage{amsmath, amssymb}
\usepackage[margin=1in]{geometry}
\usepackage{xurl}
\usepackage[citestyle=authoryear,natbib=true,backend=biber,style=apa]{biblatex}
\usepackage{graphicx}
\usepackage{lineno}
\usepackage{setspace}
\usepackage{enumerate}
\usepackage{multirow}
\usepackage{multicol}
\usepackage{booktabs}
\usepackage{siunitx}
\bibliography{lit.bib}

\begin{document}

\begin{titlepage}
    \begin{center}
        \vspace*{1cm}
        \large
        \textbf{Simplifying small area estimation with rFIA: a demonstration of tools and techniques}\\
         \normalsize
           \vspace{5mm}
         Hunter Stanke\textsuperscript{1,2}, Andrew O. Finley\textsuperscript{1}, and Grant M. Domke\textsuperscript{3}\\
         \vspace{5mm}
    
    \end{center}
    \small
    \textsuperscript{1}Department of Forestry, Michigan State University, East Lansing, MI, USA\\
    \textsuperscript{2}School of Environmental and Forest Sciences, University of Washington, Seattle, WA, USA\\
    \textsuperscript{3}Forest Service, Northern Research Station, US Department of Agriculture, St Paul, MN, USA\\ 
    \noindent \textbf{Corresponding Author}: Hunter Stanke, telephone: (269) 221-4745; email: stankehu@msu.edu

    \begin{abstract}
    \noindent The United States (US) Department of Agriculture Forest Service Forest Inventory and Analysis (FIA) program operates the national forest inventory of the US. Traditionally, the FIA program has relied on sample-based approaches---permanent plot networks and associated design-based estimators---to estimate forest variables across large geographic areas and long periods of time. These approaches generally offer unbiased inference on large domains but fail to provide reliable estimates for small domains due to low sample sizes. Rising demand for small domain estimates will thus require the FIA program to adopt non-traditional estimation approaches that are capable of delivering defensible estimates of forest variables at increased spatial and temporal resolution, without the expense of collecting additional field data. In light of this challenge, the development of small area estimation (SAE) methods---estimation techniques that support inference on small domains---for FIA data has become an active and highly productive area of research. Yet, SAE methods remain difficult to apply to FIA data, due in part to the complex data structures and survey design used by the FIA program. Herein, we present the potential of rFIA, an open-source R package designed to increase the accessibility of FIA data, to simplify the application of a broad suite of SAE methods to FIA data. We demonstrate this potential via two case studies: (1) estimation of contemporary county-level forest carbon stocks across the conterminous US using a spatial Fay-Herriot model; and (2) temporally-explicit estimation of multi-decadal trends in merchantable wood volume in Washington County, Maine using a Bayesian multi-level model. In both cases, we show the application of SAE techniques offers considerable improvements in precision over FIA's traditional, post-stratified estimators. Finally, we offer a discussion of the potential role that rFIA and other open-source tools might play in accelerating the adoption of SAE techniques among users of FIA data.

    \end{abstract}

\end{titlepage}

\section*{Introduction}

The United States (US) Department of Agriculture Forest Inventory and Analysis (FIA) program conducts the US national forest inventory (NFI), collecting data describing the condition of forest ecosystems on a large network of permanent inventory plots distributed across all lands in the nation \citep{smith2002forest}. These data offer a unique and powerful resource for determining the extent, magnitude, and causes of long-term changes in forest health, timber resources, and forest landowner characteristics across large regions in the US \citep{wurtzebach2020supporting}. The FIA program has traditionally relied on post-stratification to improve precision of point and change estimates \citep{westfall2011post, bechtold2005enhanced}. Like other NFIs \citep{breidenbach2012small, kohl2006sampling}, FIA has experienced increased demand for estimates within smaller spatial, temporal, and biophysical domains than post-stratification can reasonably deliver (e.g., annual, stand-level estimates). The development of estimation techniques that support inference on small domains---referred to as small area estimation (SAE) methods---using FIA data is an active area of research, with considerable progress made in the last decade \citep{hou2021updating, coulston2021enhancing, schroeder2014improving, lister2020use}. SAE methods are numerous and diverse, though most seek to improve inference on small domains by making use of statistical models and auxiliary information that is correlated with target variables \citep{rao2015small}.

Despite recent progress in SAE method development, many FIA data users are likely to find such techniques difficult to implement due to limitations in data accessibility and complexity in survey design. Here, we demonstrate the potential of rFIA \citep{stanke2020rfia}, an open-source R package \citep{r2021}, to reduce barriers in data access that arise from complexity in data coding, database structure, and Structured Query Language used by the FIA program. Using a simple yet powerful design, rFIA implements the post-stratified, design-based estimation procedures described in \citet{bechtold2005enhanced} for over 60 forest variables and allows users to return intermediate summaries of all variables for use in modeling studies (i.e., plot, condition, and/or tree-level). Further, target variables can be easily estimated for domains defined by any combination of spatial zones (i.e., spatial polygons), temporal extents (e.g., most recent measurements), and/or biophysical attributes (e.g., species, site classifications). 

Model-based SAE techniques offer a valuable alternative to the design-based, post-stratified estimators implemented in rFIA. Model-based SAE methods often seek to borrow information from non-target domains (e.g., from neighboring spatial zones if domains are defined by spatial boundaries) and auxiliary data (e.g., remote sensing data) to improve precision of estimated quantities for a domain of interest, and can generally be classified into two distinct groups: unit-level and domain-level (also referred to as area-level) models. Unit-level models are constructed at the level of population units, where population units are defined as the minimal units that can be sampled from a target population. With respect to FIA's survey design, field plots represent population units (in the finite population sense) and target populations are defined by any spatial and/or temporal region with known extent. Unit-level models relate target variables measured on sampled population units to auxiliary data that is available for all population units (e.g., wall-to-wall remote sensing data) in order to predict quantities of the target variables for a domain of interest (i.e., where domains are defined by some combination of population units) \citep{rao2015small}. In contrast, domain-level models are constructed at the level of domains. Here, domain-specific auxiliary information (e.g., county-level census data, where counties represent domains) is related to post-stratified or direct estimates within corresponding domains \citep{rao2015small}. Hence, domain-level models effectively ``adjust'' direct domain estimates in light of auxiliary information. 

By design, rFIA does not implement model-based SAE techniques directly, owing to their exceptional variety and requirements for thorough model checking and validation. Rather, rFIA automates the process of summarizing FIA data to a form that is appropriate for input to a wide variety of unit- and domain-level SAE models. Hence, rFIA allows the user to focus their attention on model development and data output, as opposed to the intricacies of FIA's data structure and sampling design. 

Here we present two case studies chosen to demonstrate some aspects of rFIA's potential to simplify model-based SAE applications using FIA data. First, we use the post-stratified estimators implemented in rFIA to estimate current forest carbon stocks within counties across the conterminous US (CONUS), and develop a domain-level spatial Fay-Herriot SAE model to couple these direct estimates with auxiliary climate variables and improve precision of estimated carbon stocks. Second, we derive a temporally-explicit unit-level estimator of total merchantable volume for a small spatial domain in Maine (i.e., Washington County), and compare precision of the model-based estimator to that of a design-based, post-stratified estimator of merchantable volume for the domains of interest (Washington County, all years over the period 1999-2025). Specifically, we use rFIA to extract survey design information associated with current volume inventories in the State, and produce plot-level summaries of merchantable volume for all plot visits since 1999. We then develop a Bayesian multi-level model to estimate merchantable volume at annual time-steps, and use the approach presented in \citet{little2004model} to derive a robust model-based estimator of total merchantable volume for all domains of interest. All code and data used in these case studies are available in Appendices A and B, on GitHub (\url{https://github.com/hunter-stanke/FGC_rFIA_SAE}), and at our official website (\url{https://rfia.netlify.app}).

\section*{Methods}

\subsection*{FIA data}

\subsubsection*{Data collection}
Since 1999 FIA has operated an extensive nationally-consistent annual forest survey designed to monitor changes in forests across all lands in the US \citep{smith2002forest}. The program measures forest variables on a network of permanent ground plots that are systematically distributed at a base intensity of approximately 1 plot per 2428 hectares across the US \citep{smith2002forest}. Data collected on ground plots are stored in a large, public database (i.e., the FIA Database), however the true locations of ground plots are not released in order to protect the ecological integrity of plots and the privacy rights of private landowners \citep{shaw2008benefits}.

For trees 12.7cm diameter at breast height (d.b.h.) and larger, tree attributes (e.g., species, live/dead, mortality agent) and variables (e.g., d.b.h., height, volume) are measured on a cluster of four 168$\mathrm{m}^2$ subplots at each plot location \citep{bechtold2005enhanced}. Trees 2.54-12.7cm d.b.h. are measured on a microplot (13.5$\mathrm{m}^2$) contained within each subplot, and rare events such as very large trees are measured on an optional macroplot (1012$\mathrm{m}^2$) surrounding each subplot \citep{bechtold2005enhanced}. Importantly, some variables in the FIA database, like tree biomass and carbon, are modeled from variables measured on field plots and auxiliary variables, such as mean annual temperature, that are joined with the plots based on their spatial location. 

\subsubsection*{Survey design}
Traditionally, the FIA program has used post-stratification to improve precision of point and change estimates, account for variability in non-response rates, and to allow sample intensity to vary across regions \citep{bechtold2005enhanced, tinkham2018applications, smith2002forest}. Importantly, post-stratification is applied to populations defined by a set of exhaustive and mutually exclusive geographic units with known areas---known as estimation units using FIA's terminology. Estimation units are often formed from administrative boundaries, for example counties, county groups, or large ownerships and are constrained by State boundaries (i.e., estimation units can only fall within one State). FIA implements post-stratification by dividing each estimation unit into relatively homogeneous strata using wall-to-wall remotely-sensed imagery. Strata are designed to minimize within-strata sample variances, while ensuring constant within-strata sample intensity. In short, FIA's survey design is hierarchical and area-based: States are comprised of multiple estimation units, estimation units are divided into multiple strata, and strata contain multiple inventory plots. We refer readers to \citet{bechtold2005enhanced} for a complete description of FIA's post-stratified survey design.

FIA uses an annual panel system to estimate current inventories and change. Inventory cycles---the period of time required to measure all ground plots with at least one forest condition within an estimation unit---are generally 5-7 years in length in the eastern US, and 10 years in length in the western US \citep{bechtold2005enhanced}. A mutually exclusive and spatially-balanced subset of ground plots with at least one forest condition are measured in each year of an inventory cycle, forming a series of independent annual panels. For example in an ideal 5-year inventory cycle, 20\% of ground plots are measured annually, such that 100\% of plots are measured once between Year 1 and Year 5. In Year 6, the subset of plots measured in Year 1 are remeasured, and a second inventory cycle emerges consisting of all plots measured between Year 2 to Year 6 (not independent of the previous cycle, as 80\% of measurements are shared).

Precision of point and change estimates can often be improved by combining annual panels within an inventory cycle (i.e., by augmenting current data with data collected previously). While FIA does not prescribe a core procedure for combining panels \citep{bechtold2005enhanced}, the temporally-indifferent approach, which effectively pools data from annual panels into a single periodic inventory, is the most widely known and used. From our example 5-year inventory cycle above, the temporally-indifferent approach pools all data collected between Years 1-5 and computes point estimates from the aggregated sample, assuming all plots are measured simultaneously at the end of the inventory cycle. Estimates of change could first be computed in Year 6 in our example (consisting of a single annual panel, 20\% of remeasured plots), and change estimates for a full inventory cycle could first be computed following Year 10. In the case studies that follow, we use the periodic, or temporally-indifferent, approach to estimate contemporary carbon stocks across the CONUS, and the post-stratified estimator applied to individual annual panels to characterize temporal trends in merchantable volume in Maine. Importantly, both approaches rely on the same direct post-stratified estimator, differing only in their treatment of time as dimension of the survey design (i.e., the temporal subset of data that the estimators are applied to). 

\subsection*{The rFIA R package}
rFIA is an open source package for the statistical computing environment R \citep{r2021}, and was designed to simplify the process of working with FIA data. Specifically, rFIA alleviates hurdles arising from FIA's complex survey design and database structure by offering a simple and highly flexible toolset for data acquisition and management (e.g., downloading and storing FIA data), population estimation (e.g., estimation of totals and ratios for domains of interest), and alternative summary of FIA data (e.g., plot-level summaries of forest variables). We provide a brief description of the key features of rFIA here, and refer readers to \cite{stanke2020rfia} for a detailed description of the package and our official website (\url{https://rfia.netlify.app/}) for example code and details regarding package installation. 

Core functions in the rFIA R package can be divided into three categories: (1) utility functions designed to acquire, load, and save modifications to FIA data; (2) subset functions designed to help users navigate FIA's survey design and subset inventories of interest in their applications; and (3) estimator functions that ingest raw FIA data and produce population estimates (e.g., totals, ratios, and associated variances) or intermediate-level summaries (e.g., plot- or tree-level summaries) of forest variables within user-defined populations of interest. Table \ref{tab:core_funtions} provides a brief description of the rFIA functions used in the case studies presented herein, and Appendices A and B provide all associated code required to reproduce these case studies.  

\begin{table}[]
    \centering
    \begin{tabular}{l l}
        \toprule
         rFIA Function & Description \vspace{0.25 em} \\
        \midrule 
         \textit{Utility Functions} \\
         \hskip 2em getFIA() & Download FIA data, load into R, and optionally save to disk \\
         \hskip 2em readFIA() & Load FIA database into R environment from disk \\
         \textit{Subset Functions} \\
         \hskip 2em clipFIA() & Spatial and temporal queries for FIA data \\
         \hskip 2em getDesignInfo() & Extract survey design information for post-stratified inventories \vspace{0.25 em} \\
        \textit{Estimator Functions} \\
         \hskip 2em carbon() & Estimate carbon stocks by IPCC forest carbon pools  \\
         \hskip 2em volume() & Estimate merchantable volume on standing trees \\
        \bottomrule
    \end{tabular}
    \caption{Descriptions of core rFIA functions used in case studies presented herein.}
    \label{tab:core_funtions}
\end{table}

By default, rFIA implements standard estimation routines used by the FIA program---post-stratified estimators and a temporally-indifferent (i.e., periodic) approach to combining annual panels within inventory cycles---to produce population estimates for more than 60 forest variables. These estimation routines have been tested extensively across Forest Service regions and potential domains of interest (e.g., defined by species, land types) to ensure national-consistency and appropriate behavior of the estimators under a broad range of user-inputs. Furthermore, resulting estimates have been validated against official FIA estimation tools (i.e., EVALIDator) \citep{evalidator}, and found to be accurate to two decimal places for all forest variables \citep{stanke2020rfia}. In addition to standard estimation approaches, rFIA offers users the ability to produce population estimates for individual annual panels or combine annual panels within an inventory cycle using a moving-average approach with potentially time-decaying weights (simple, linear, or exponential moving averages). We refer readers to Section 2.2 of \cite{stanke2020rfia} for additional details on the estimation routines implemented in rFIA.

\subsection*{Domain-level model for forest carbon stocks}
\begin{figure}[t!]
    \centering
    \includegraphics[width=6in]{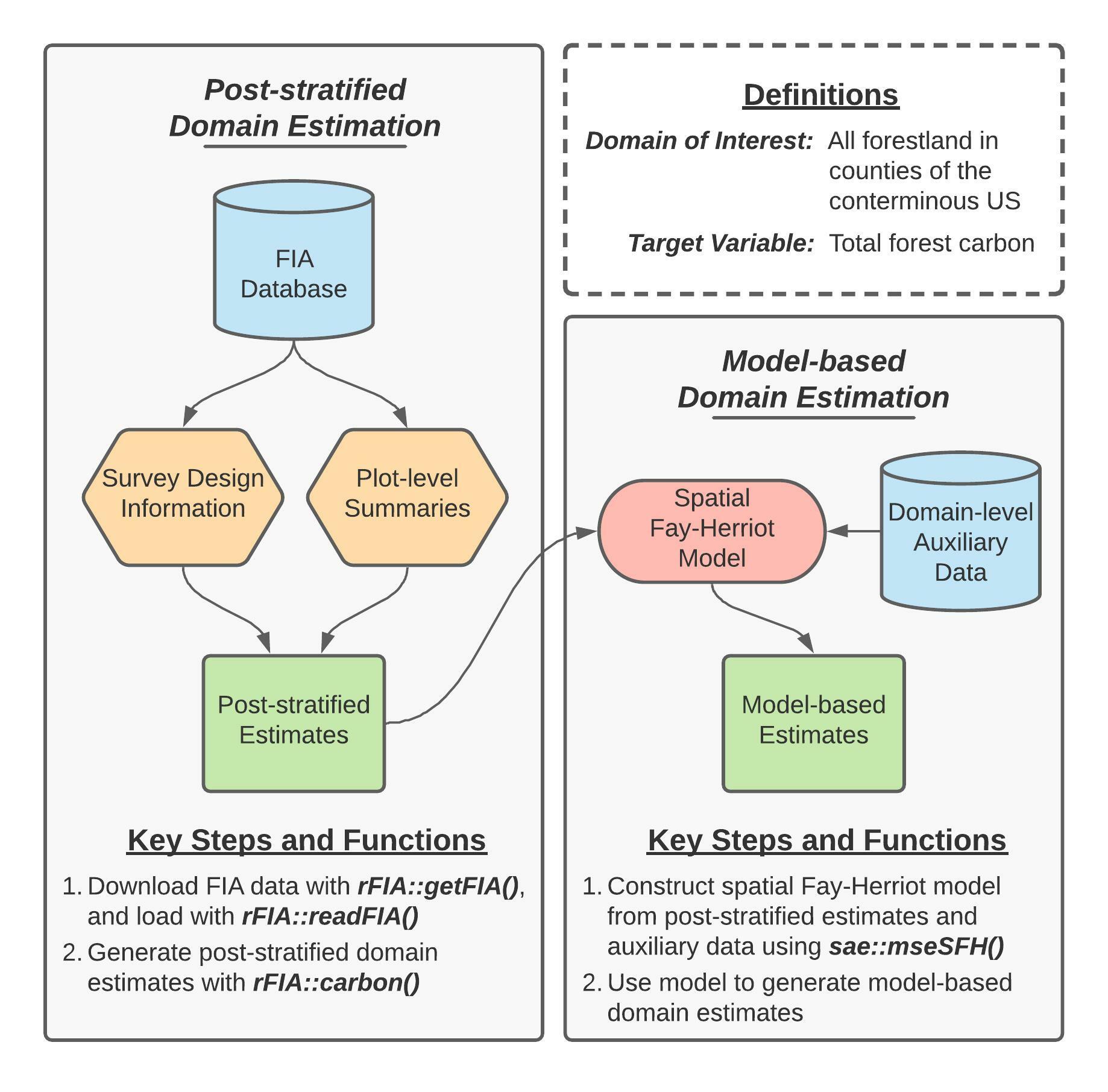}
    \caption{Concept map illustrating key steps, functions, and workflows used in the development of our spatial Fay-Herriot model of forest carbon stocks in conterminous US. Here, blue cylinders represent data inputs, orange hexagons represent intermediate data products, red ovals represent models, and green rectangles represent domain estimates.}
    \label{fig:diagram1}
\end{figure}

To demonstrate rFIA's capacity to simplify development of domain-level small area estimators, we estimate contemporary forest carbon stocks by county across the CONUS using a spatial Fay-Herriot model \citep{fay1979estimates, petrucci2006small}. This process consists of two primary stages: (1) produce post-stratified estimates of carbon stocks and associated variances for all forestland in each county (i.e., domain), and (2) ``smooth'' post-stratified estimates using a model constructed from domain-average climate variables and spatial random effects to improve precision of estimated quantities within each domain. Figure \ref{fig:diagram1} provides a conceptual diagram that illustrates key steps in our general estimation approach.

FIA measures/models forest carbon variables on all forested portions of inventory plots \citep{domke2021}. Here, forestland is defined as land with at least 10\% tree canopy cover (or had previously, or is expected to have in the future) that occurs in a patch of at least 0.4 ha in extent and that is not narrower than 37 meters. The carbon() function in rFIA draws from forest carbon variables to produce population estimates of forest carbon stocks, where carbon stocks include the following ecosystem components: live overstory, live understory, standing dead wood, down dead wood, litter, and soil organic material. Here live overstory, live understory, and standing dead wood encompass both aboveground and belowground carbon stocks. 

We used rFIA to download an appropriate subset of the FIA Database from the FIA DataMart \citep{datamart}, and select the most recent subset of current volume inventories within each State across the CONUS. We then used the carbon() function to estimate total carbon stocks within counties using the periodic, temporally-indifferent approach (i.e., the same methods implemented by EVALIDator \citep{evalidator}). Here, total carbon stocks are a sum of all ecosystem components across public and private forestland, and are expressed as a population total. We convert estimates of population totals (tons CO2e) to population means (tons  $\mathrm{CO2e}  \cdot  \mathrm{ha}^{-1}$) by dividing population totals by the areal extent of each county (known quantities). Similarly, we convert the variance of the population total to the variance of the population mean by dividing by the square of the areal extent of each domain. 

We next fit a spatial Fay-Herriot model to the post-stratified estimates of population means, using the sae R package \citep{molina2015sae}. Fay-Herriot models are widely used in small area estimation and generally use domain-level auxiliary data in an attempt to improve the precision of domain estimates for a target variable. These models are often defined in two stages, in which variability arising from imperfect observation of the target variable within a domain (e.g., variability arising from sampling) is modeled separately from variability arising from functional processes (e.g., processes represented in the auxiliary data). This framework is particularly useful as it allows estimation of relationships between auxiliary variables and the true state of a target variable, without requiring that the true state of the target variable be known. Instead, the probabilistic linkage between imperfect observations of the target variable (e.g., sample-based estimates with known error) and its true state are used to estimate these relationships,  thereby allowing information to be ``borrowed'' across domains (e.g., via shared regression coefficients) and often improving the precision of domain estimates for the target variable \citep{molina2015sae}.

Let $\bar{Y}_d$ denote the estimated population mean of county $d$ obtained via the post-stratified estimators from rFIA, and $\mathrm{v}( \bar{Y}_{d} )$ the estimated variance of $\bar{Y}_d$. Importantly, the estimators of $\bar{Y}_d$ and $\mathrm{v}( \bar{Y}_{d} )$ are derived under a design-based framework, and hence can be assumed unbiased for large samples (an assumption that is potentially violated for domains with few observations). The spatial Fay-Herriot model for county $d$ in $1,2, \ldots, D$, where $D$ is the number of counties ($D=3107$), is then defined as
\begin{linenomath*}
\label{model1}
\begin{align}
   \bar{Y}_d &= Z_{d} + \epsilon_{d}, \label{model1a} \\ 
   Z_{d} &= \mathbf{x}_{d}^\top \boldsymbol{\beta} + v_{d}, \label{model1b}
\end{align}
\end{linenomath*}
where $Z_d$ denotes the true, but unobserved value of the population mean in county $d$, and $\epsilon_{d}$ is a normally distributed error term with zero mean and variance $\mathrm{v}( \bar{Y}_{d} )$. Eq \ref{model1a} represents post-stratified estimates of county-level population means from rFIA as imperfect observations of true (unobserved) county-level population means. In other words, we represent the post-stratified estimate for domain $d$ as being drawn from a normal distribution with mean $Z_{d}$ (unobserved, and to be estimated) and variance $\mathrm{v}( \bar{Y}_{d} )$ (estimated directly from FIA data, assumed to be known). 

In Eq \ref{model1b}, $\mathbf{x}_d$ is a vector of length three comprising an intercept and two climate predictors for county $d$, and $\boldsymbol{\beta}$ is an associated vector of regression coefficients. Climate predictors include mean annual temperature and precipitation, and were obtained from the long-term (30-year) climate normals hosted in the PRISM climate dataset \citep{prism}. Climate normals were distributed on a 800$\mathrm{m}^2$ grid spanning the CONUS, and we took an average of grid cells within each county to produce domain-level climate predictors. The collection of county random effects  $\mathbf{v} = (v_1, v_2, \ldots, v_D)^\top$ is assumed to follow a first order simultaneous autoregressive (SAR) process
\begin{linenomath*}
\begin{equation}
   \mathbf{v} = \rho \mathbf{W} \mathbf{v} + \boldsymbol{\tau}, \label{model1c}
\end{equation}
\end{linenomath*}
where $\rho$ is the autocorrelation parameter defined on the range (-1, 1), and each element of the vector $\boldsymbol{\tau}$ is a normally distributed error term with mean zero and variance $\sigma_v^2$. Finally, $\mathbf{W}$ is a $D \times D$ row-standardized county proximity matrix. In words, Eqs \ref{model1b}-\ref{model1c} represent the true county-level population means ($Z_{d}$, unobserved) as a linear function of our climate predictors and a first-order spatial process which accounts for all variation in $Z_{d}$ unexplained by $\mathbf{x}_{d}^\top \boldsymbol{\beta}$ (i.e., linear relationship between population means and climate variables).

\citet{petrucci2006small} present an empirical best linear unbiased predictor (EBLUP) under the Fay-Herriot model with spatially correlated random effects, and an analytic estimator of the mean squared error (MSE) of the EBLUP is described in \citet{singh2005spatio}. We use the sae R package \citep{molina2015sae} to fit the model described in Eqs \ref{model1a}-\ref{model1c}, and obtain the EBLUP of population means $\bar{Y}_{d}^{\mathrm{EBLUP}}$ and associated mean squared error $\mathrm{MSE}(\bar{Y}_{d}^{\mathrm{EBLUP}})$ for all domains via restricted maximum likelihood. 

We use the relative standard error (RSE, expressed as a percentage) as a standardized measure of precision of the estimators of forest carbon stocks
\begin{linenomath*}
\begin{align}
   \mathrm{RSE}_{d}^{PS} &= \frac{100 \, [\mathrm{v}( \bar{Y}_{d} )]^{0.5}}{\bar{Y}_{d}}, \\
   \mathrm{RSE}_{d}^{EBLUP} &= \frac{100 \,[\mathrm{MSE}(\bar{Y}_{d}^{\mathrm{EBLUP}})]^{0.5}}{\bar{Y}_{d}^{\mathrm{EBLUP}}}.
\end{align}
\end{linenomath*}
Here, a lower RSE indicates higher precision. Following \citet{coulston2021enhancing}, we compare the precision of post-stratified (design-based) and model-based estimators of forest carbon stocks using the ratio of their respective standard errors for each domain
\begin{linenomath*}
\begin{align}
   \mathrm{SER}_{d} &= \frac{[\mathrm{MSE}(\bar{Y}_{d}^{\mathrm{EBLUP}})]^{0.5}}{[\mathrm{v}( \bar{Y}_{d} )]^{0.5}},
\end{align}
\end{linenomath*}
where SER$_{d}$ denotes the ratio of the standard error of the post-stratified estimator (assumed unbiased) to that of the EBLUP for domain $d$ (derived from MSE, cannot be assured to be unbiased). Hence, a SER less than one indicates the EBLUP yields more precise estimates of forest carbon stocks than the post-stratified estimator.

\subsection*{Unit-level model for merchantable wood volume trends}
To demonstrate rFIA's capacity to simplify development of unit-level small domain estimators of forest variables, we use a Bayesian multi-level model to estimate multi-decadal trends in merchantable wood volume in Washington County, Maine. This process consists of four primary stages: (1) extract survey design information associated with the most recent ``current volume'' inventory in Maine; (2) produce plot-level summaries of merchantable volume for all FIA plot visits within our target population; (3) fit a Bayesian multi-level linear model to estimate plot- and stratum-level trends in mean merchantable volume, accounting for repeated inventory plot observations; and (4) summarize regression model coefficients using post-stratified design weights, yielding a robust model-based estimator of temporal trends in total merchantable wood volume across Washington County. Note that in this case study, domains are defined by spatial, temporal, and biophysical boundaries, i.e., by the spatial boundary of Washington County, by individual years over the period 1999-2025, and by the unknown extent of timberland (defined below) in the region.

\begin{figure}[t!]
    \centering
    \includegraphics[width=6in]{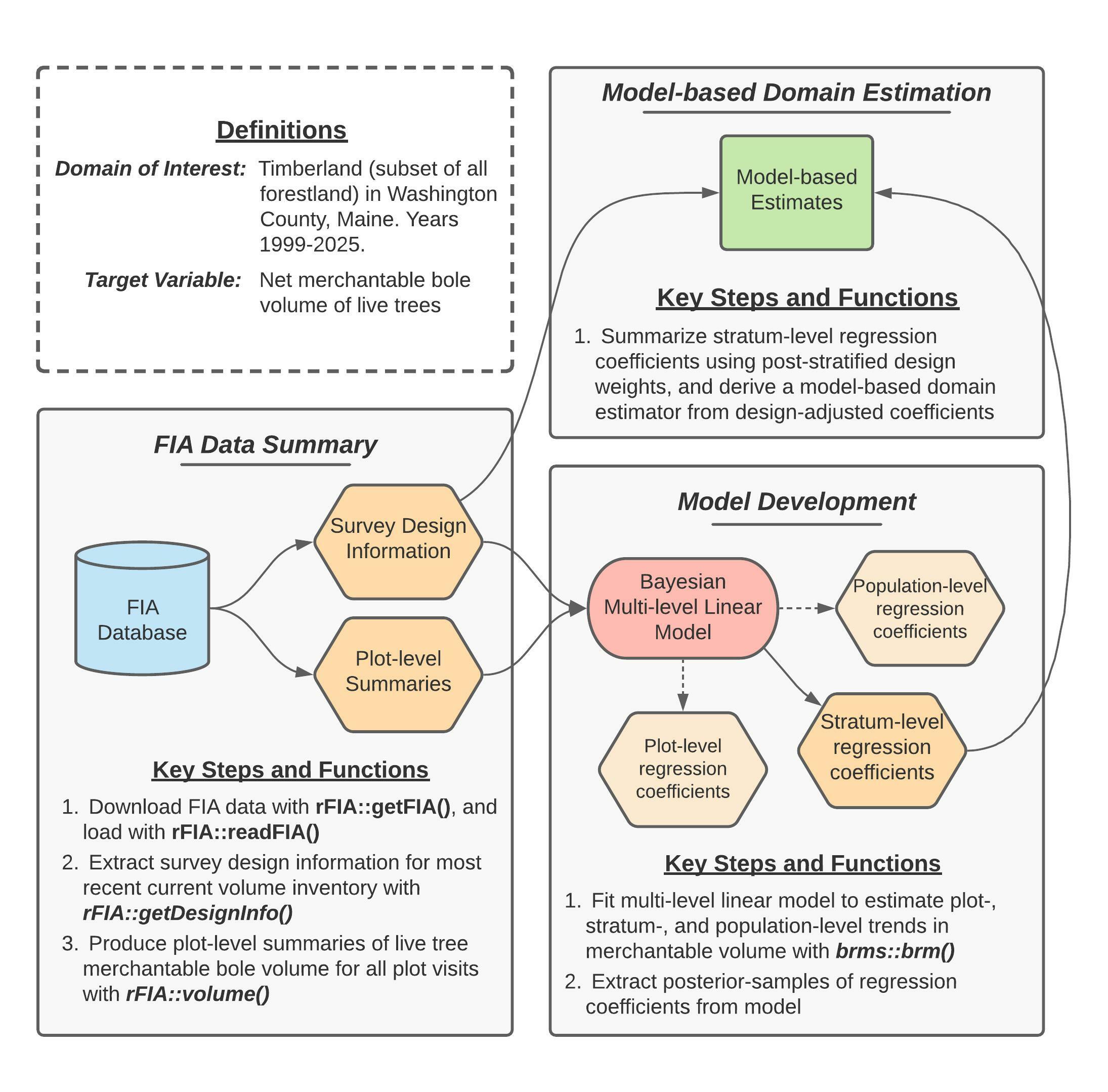}
    \caption{Concept map illustrating key steps, functions, and workflows used in to estimate multi-decadal trends in merchantable wood volume in Washington County, Maine. Here, blue cylinders represent data inputs, orange hexagons represent intermediate data products, red ovals represent models, and green rectangles represent domain estimates.}
    \label{fig:diagram2}
\end{figure}

FIA records merchantable wood volume of all trees (d.b.h.$\, \geq \,$12.7cm) on forested inventory plots. The volume() function in rFIA uses these observations to produce population estimates and plot-level summaries of merchantable wood volume in the bole and sawtimber portions of trees. We consider net merchantable bole volume herein, defined as the volume of wood in the central stem of trees (d.b.h.$\, \geq \,$12.7cm), from a 30.5cm stump to a minimum 10.2cm top diameter, or to where the central stem breaks into limbs all of which are $\leq \,$10.2cm in diameter \citep{database}. Volume loss due to rot and form defect are deducted. Further, FIA defines timberland as the subset of forestland that is capable of producing crops of industrial wood and is not withdrawn from timber utilization by legal statute or administrative regulation (i.e., it excludes wilderness areas) \citep{database}. 

We used rFIA to download the Maine subset of the FIA Database from the FIA DataMart \citep{datamart}, extract survey design information (i.e., stratum and population areas) for the most recent current volume inventory in the State (2019 inventory), and summarize plot-level net merchantable bole volume for all plot-visits in the State since the onset of the annual FIA program (i.e., first plots measured in 1999). Here, plot-level summaries of merchantable volume are simply a sum of merchantable volume on all trees within our domain of interest---timberland in Washington County---at each inventory plot, expressed on a per-area basis ($\mathrm{m}^{3} \cdot \mathrm{ha}^{-1}$). All plots outside our domain of interest (e.g., non-forested) receive a value of zero. 

In the 2019 inventory, Washington County is split into three distinct estimation units (split into private and public ownerships, and inland census water). As FIA's estimation units are geographically distinct (i.e., independent populations), we combine these estimation units into a single target population representing Washington County. Importantly, FIA's estimation units should not be confused with population units in a finite sampling framework. Estimation units can be seen as minimum target populations for estimation using FIA's survey design. These populations are comprised of many population units, some of which may be sampled (i.e., plot locations). 

We next formulate a multi-level linear model to characterize plot-, stratum-, and domain-level trends in merchantable wood volume from our visit-level summaries. By explicitly acknowledging the nested, hierarchical nature of FIA's survey design in our multi-level model, we can derive inference at multiple scales simultaneously (e.g., estimation of both plot- and stratum-level trends), partition estimated variance (i.e., uncertainty) across scales, and improve parameter estimates by allowing partial-pooling of information within groups (e.g., when few observations are available on a plot, estimated trends are ``pulled'' towards the stratum-level mean). This is in contrast to conventional approaches that may perform independent linear regressions for each plot (i.e., no pooling of information) or combine data from all plots within a stratum and perform a single linear regression (i.e., complete pooling of information) to estimate trends across scales. 

Let $y_{hij}$ denote the merchantable bole volume within our domain of interest that was observed at visit $j$, on plot $i$, belonging to stratum $h$. Further, let $t_{hij}$ denote the year of visit $j$ on plot $i$, relative to onset of the annual FIA program (i.e., $t=0,1,2$ for plots visited in 1999, 2000, 2001, etc.). Our model is then defined as
\begin{linenomath*}
\begin{equation}\label{model2}
    y_{hij} = \alpha_{hi} + \beta_{hi} \cdot t_{hij} + \epsilon_{hij},
\end{equation}
\end{linenomath*}
where $\alpha_{hi}$ is a plot-level intercept term describing the mean merchantable volume at plot $i$, belonging to stratum $h$, in 1999 (i.e., onset of the annual FIA program, $t=0$), and $\beta_{hi}$ is a plot-level slope term describing the average annual change in mean merchantable volume at plot $i$, belonging to stratum $h$, over the period 1999-2019. The error term $\epsilon_{hij}$ is assumed normally-distributed with zero mean and constant variance. 

Trends in merchantable volume are expected to vary both among plots (e.g., growth rates vary by forest type, and some plots may be harvested) and among strata (e.g., predominately forested vs non-forested strata). We model this variability by treating plot-level parameters ($\alpha_{hi}$ and $\beta_{hi}$) as random effects that follow distributions defined by associated stratum-level parameters ($\alpha_{h}$ and $\beta_{h}$), and similarly treating stratum-level parameters as random effects that follow distributions defined by a set of population-level parameters ($\alpha$ and $\beta$):
\begin{linenomath*}
\begin{align}
    \alpha_{hi} &\sim \mathrm{normal}(\alpha_{h}, \sigma_{\alpha_{hi}}^2), \label{model2:ahi} \\
    \beta_{hi} &\sim \mathrm{normal}(\beta_{h}, \sigma_{\beta_{hi}}^2), \label{model2:bhi} \\
    \alpha_{h} &\sim \mathrm{normal}(\alpha, \sigma_{\alpha_{h}}^2), \label{model2:ah} \\ 
    \beta_{h} &\sim \mathrm{normal}(\beta, \sigma_{\beta_{h}}^2), \label{model2:bh}
\end{align}
\end{linenomath*}
where $\sigma_{\alpha_{hi}}^2$ and $\sigma_{\beta_{hi}}^2$ are the stratum-level (among plot) variances of the regression coefficients, and $\sigma_{\alpha_{h}}^2$ and $\sigma_{\beta_{h}}^2$ are the associated population-level (among stratum) variances. In words, Eq 7 states that plot-level trends (defined by $\alpha_{hi}$ and $\beta_{hi}$, for each plot in $i=\{1, \ldots, 310\}$) are estimated from data collected at each visit of an FIA plot ($y_{hij}$), Eqs 8-9 state that stratum-level trends (defined by $\alpha_{h}$ and $\beta_{h}$, defined for each stratum in $h=\{1, \ldots, 6\}$) represent an ``average'' of plot-level trends for all plots within a particular stratum, and Eqs 10-11 state that the domain-level trend represents an ``average'' of overall population-level trends. 

To complete the Bayesian specification of Eq \ref{model2} we assigned prior distributions to all parameters.  We choose weakly informative normal priors for $\alpha$ (i.e., mean 50, standard deviation 250) and $\beta$ (i.e., mean 0, standard deviation 100), and weakly informative half student-t priors for all variance terms (i.e., mean 0, scale 100, 3 degrees of freedom) \citep{gelman2006prior}. The mean and standard deviation assigned to priors for $\alpha$ and $\beta$ differ, as $\alpha$ represents a point-in-time estimate while $\beta$ represents an estimate of average annual change. Hence, assigning a prior to $\alpha$ with a positive mean reflects our knowledge of the non-negativity of the target variable (ideally would be addressed via specification of a non-negative likelihood function, but is not here due to computational constraints), and assigning a prior with zero mean to $\beta$ represents an assumption of no change in the population over time. Further, as $\beta$ represents an annual rate, we expect it's absolute value to be considerably less than the population total at a point-in-time (e.g., $\alpha$), and our assignment of a lower standard deviation to the prior on $\beta$ reflects this belief. Using these priors, we estimated the model using Hamiltonian Monte Carlo (HMC) algorithms implemented in the probabilistic programming language, Stan \citep{carpenter2017stan}, and affiliated R package, brms \citep{burkner2017brms}. We simulated three Markov chains, for a total of 4000 iterations per chain. We assessed convergence via visual inspection of traceplots, and ensured proper model specification via posterior predictive checks. 

While the set of parameters estimated in Eqs \ref{model2:ah}-\ref{model2:bh} ($\alpha$ and $\beta$) allow us to derive an estimator of population-level trends in merchantable volume, such estimators ignore variation in the size of strata (i.e., which is known from FIA's survey design) and thus may be biased towards stratum that contain a large number of plots (as sampling intensity may vary across strata, a constant relationship between plot number and stratum size cannot be assumed). This bias may be addressed, however, by adjusting population-level parameters using a product of model- and design-weights \citep{little2004model}. Let $\alpha_{h}^{*}$ and $\beta_{h}^{*}$ denote a set of posterior samples of stratum-level regression coefficients observed at a single iteration of the HMC algorithm. We then compute design-adjusted estimates of population-level regression coefficients, denoted as $\hat{\alpha}^{*}$ and $\hat{\beta}^{*}$, for each set of posterior samples as
\begin{linenomath*}
\begin{align}
    \hat{\alpha}^{*} = A^{-1} \sum_{h=1}^{H} A_h \cdot \alpha_{h}^{*}, \label{model2:a} \\
    \hat{\beta}^{*} =  A^{-1} \sum_{h=1}^{H} A_h \cdot \beta_{h}^{*}, \label{model2:b}
\end{align}
\end{linenomath*}
where $A_h$ is the known area of stratum $h$, and $A$ is the combined area of all $H$ strata (i.e., $A=\sum_{h=1}^H A_h$, equivalent to the combined area of estimation units). Here, model-weights are implicit in estimates of stratum-level parameters, arising from the hierarchical nature of the model described in Eqs \ref{model2:ahi} - \ref{model2:bhi}. In contrast, design weights are explicit, with large strata receiving more weight than small strata. In essence, we take an area-weighted mean of regression coefficients across strata to estimate trends at the population-level, thereby explicitly acknowledging features of FIA's survey design in the construction of our model-based estimator of population parameters. 

Using our adjusted population-level regression coefficients, we derive a robust model-based estimator \citep{little2004model} of the population mean and total for our domains of interest, denoted as $\bar{Y}(t)^{*}$ and $\acute{Y}(t)^{*}$, respectively:
\begin{linenomath*}
\begin{align}
    \bar{Y}(t)^{*} &= \hat{\alpha}^{*} + \hat{\beta}^{*} \cdot t, \label{model2:mean} \\
    \acute{Y}(t)^{*} &= A \cdot \bar{Y}_{t}^{*}. \label{model2:total}
\end{align}
\end{linenomath*}
Here, variability in $\bar{Y}(t)$ and $\acute{Y}(t)$ across posterior samples reflects uncertainty in the model-based estimator of the population parameters. We produce point estimates of population parameters and their associated variances from the posterior mean and variance, and obtain 95\% interval estimates from the 2.5\% and 97.5\% percentiles of the posterior samples for each population parameter. Similarly, we compute the relative standard error for each estimator as the ratio of the posterior standard deviation to the posterior mean. 

Finally, we evaluate the performance of the model-based estimator of trends in total merchantable volume by comparing model-based population estimates to post-stratified annual estimates for the same population of interest over the period 1999-2019. All post-stratified estimates were computed using the annual approach implemented in the volume function in rFIA, and hence represent estimates of individual annual panels. We have elected to use estimates for annual panels because our domains are partially defined by individual years. A direct estimator then, by definition, should draw only from data collected within a particular year to produce domain estimates. Importantly, this approach differs from standard FIA estimation procedures, which pool data from multiple (up to 10) annual panels within an inventory cycle to generate domain estimates. 

\begin{figure}[t!]
    \centering
    \includegraphics[width=6in]{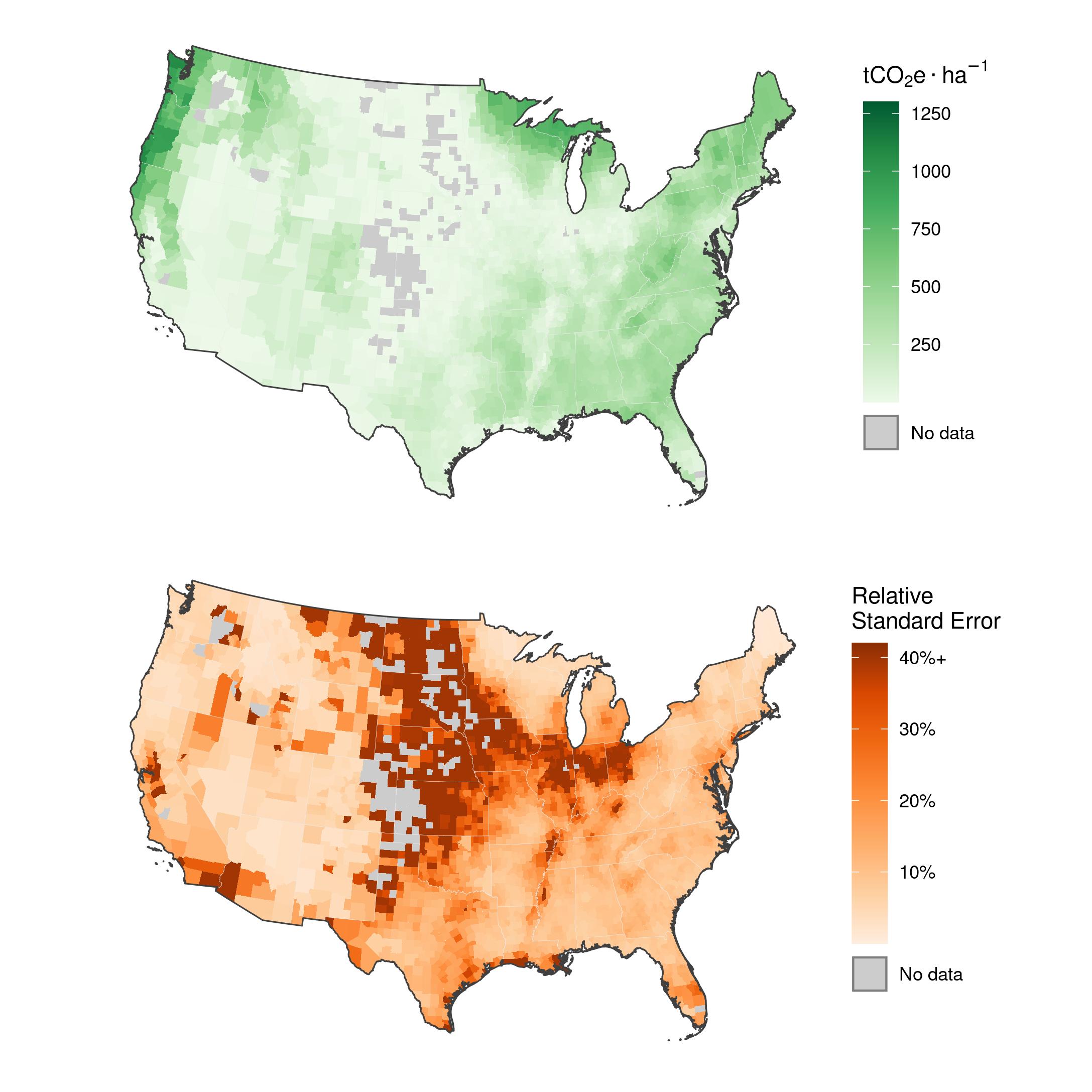}
    \caption{County-level estimates of mean forest carbon density (tons $\mathrm{CO_2}$ equivalent per hectare, $\mathrm{tCO}_2\mathrm{e} \cdot \mathrm{ha}^{-1}$) produced by the spatial Fay-Herriot model with climate predictors (top), and associated relative standard error (\%; bottom). Gray shaded counties indicate no forested FIA plots were encountered in the county during the most recent current volume inventory, i.e., post-stratified estimator of total forest carbon and associated variance for the county are equal to zero.}
    \label{fig:smoothed}
\end{figure}

\begin{figure}[t!]
    \centering
    \includegraphics[width=6in]{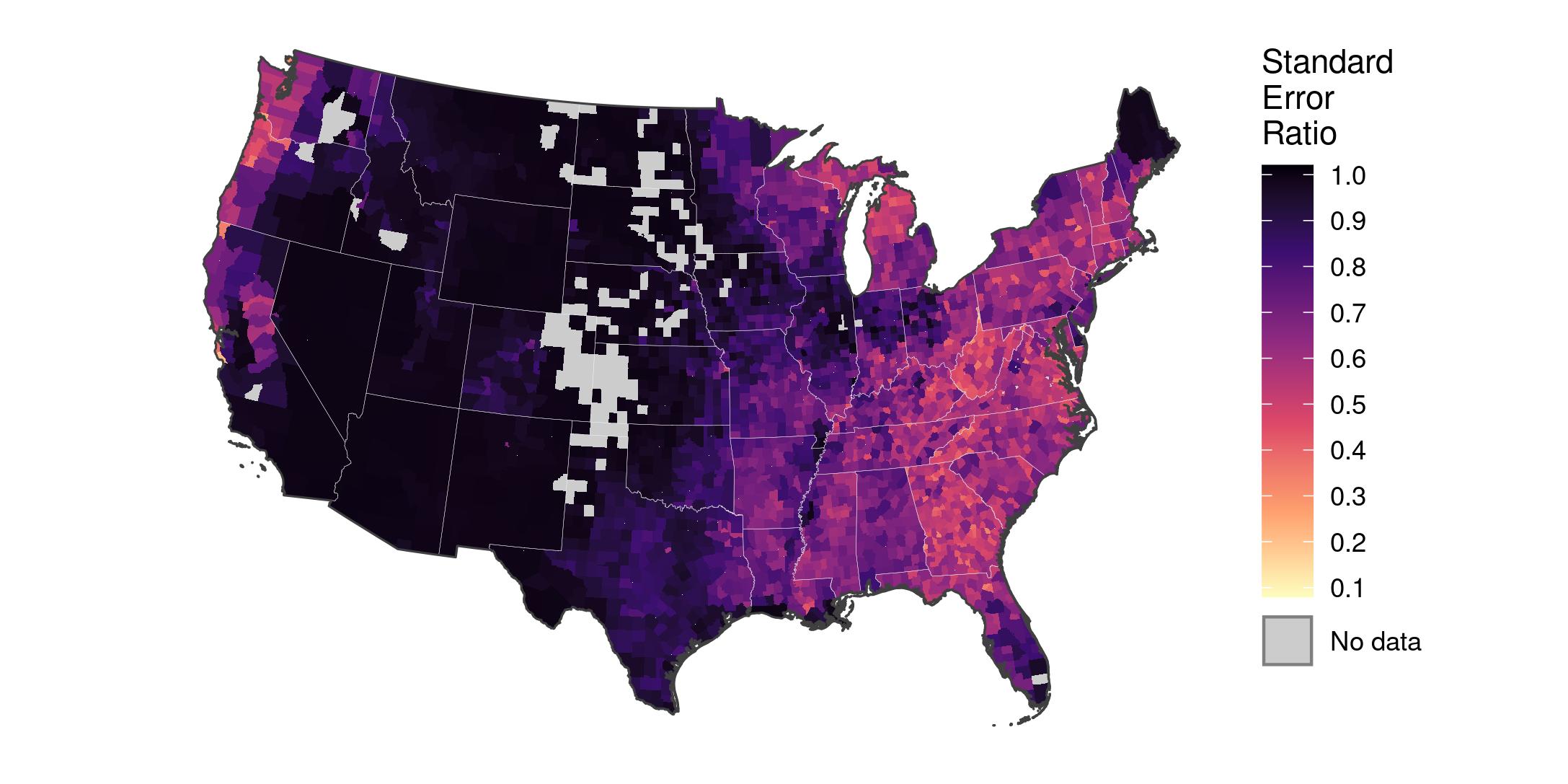}
    \caption{County-level ratios of the standard error of the spatial Fay-Herriot model-based estimator of mean forest carbon density, relative to that of the post-stratified estimator. Ratios less than one indicate the model-based approach yields a more precise estimator of forest carbon stocks than the traditional design-based approach. Gray shaded counties indicate no forested FIA plots were encountered in the county during the most recent current volume inventory, i.e., direct estimator of total forest carbon and associated variance for the county are equal to zero.}
    \label{fig:error}
\end{figure}

\begin{figure}[t!]
    \centering
    \includegraphics[width=3.5in]{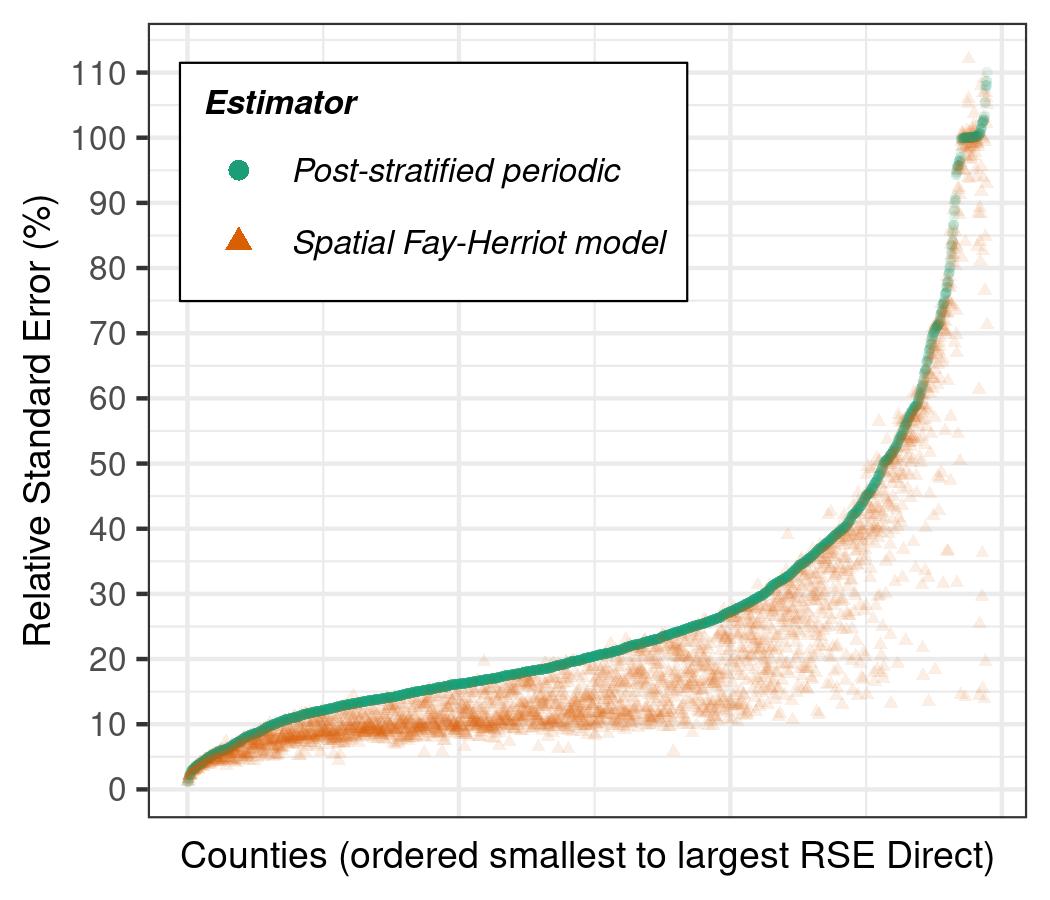}
    \caption{Relative standard error (\%) of model-based (i.e., spatial Fay-Herriot model) and post-stratified estimator estimators of mean forest carbon density by county, ordered by increasing relative standard error of the direct estimator.}
    \label{fig:carbon_cv}
\end{figure}

\begin{figure}[t!]
    \centering
    \includegraphics[width=5.5in]{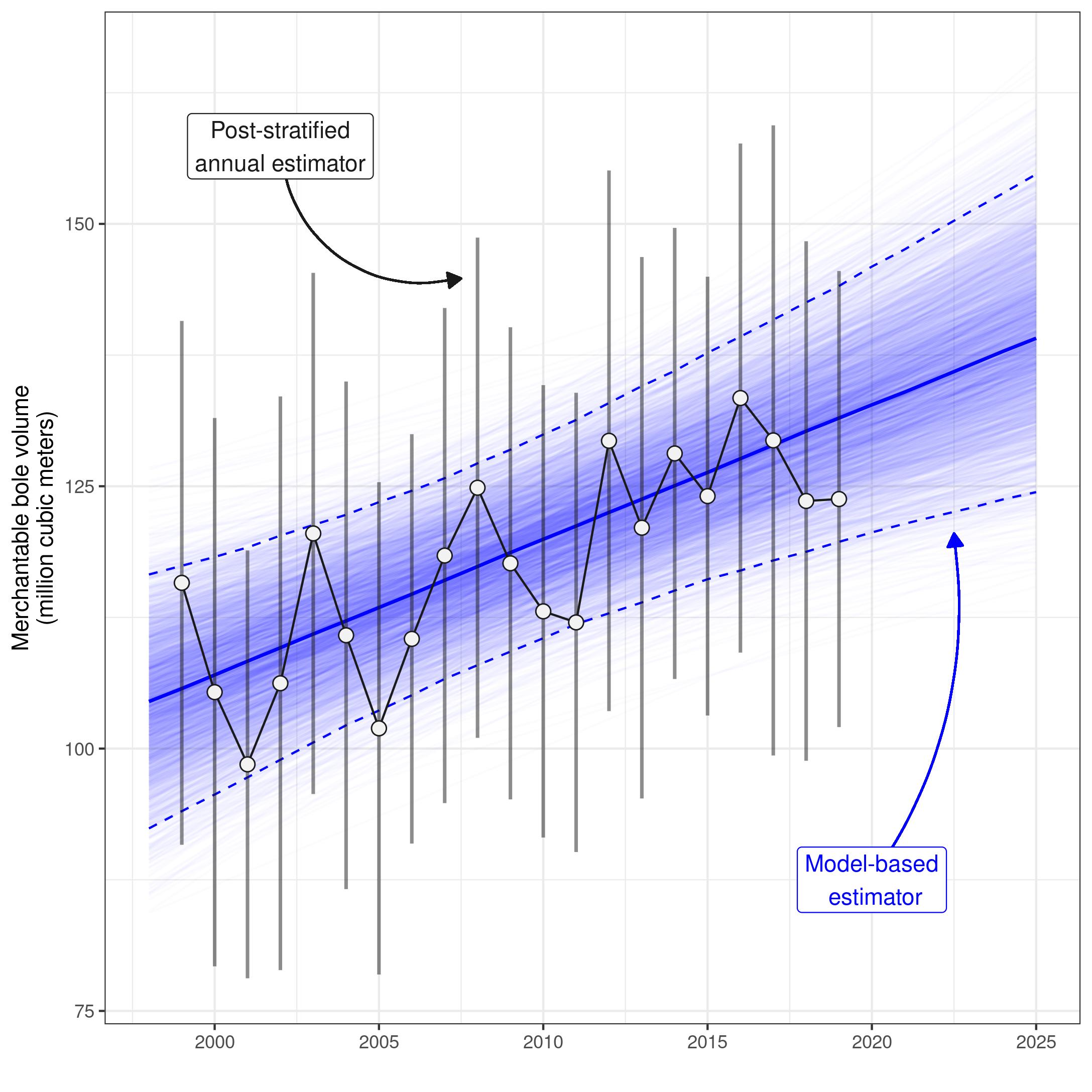}
    \caption{Annual model-based and design-based estimates of total merchantable wood volume (million $\mathrm{m}^3$) on timberland in Washington County, Maine. Model-based point estimates are derived from the posterior median of Hamiltonian Monte Carlo (HMC) samples of parameters presented in Eqs \ref{model2:mean}-\ref{model2:total}, and are represented by the solid, dark blue line. Similarly, model-based interval estimates (i.e., Bayesian 95\% credible intervals) are derived from the 2.5\% and 97.5\% quantiles of HMC samples, and are represented by dashed, dark blue lines. Further, realizations of parameters presented in Eqs \ref{model2:mean}-\ref{model2:total} from each HMC sample are represented as thin, semi-transparent blue lines. Hence the posterior predictive distribution of the model-based estimator of total merchantable wood volume can be inferred from the relative density of thin blue lines in a given region of the graph (i.e., higher density of lines indicates higher posterior probability). Annual, design-based point estimates are represented by white circles, and are connected by a solid black line. Design-based interval estimates (95\% confidence intervals) associated with each annual point estimate are presented as vertical gray bars. All design-based estimates were produced using the annual, post-stratified estimation approach implemented in rFIA \citep{stanke2020rfia}. }
    \label{fig:volume}
\end{figure}

\begin{figure}[t!]
    \centering
    \includegraphics[width=3.5in]{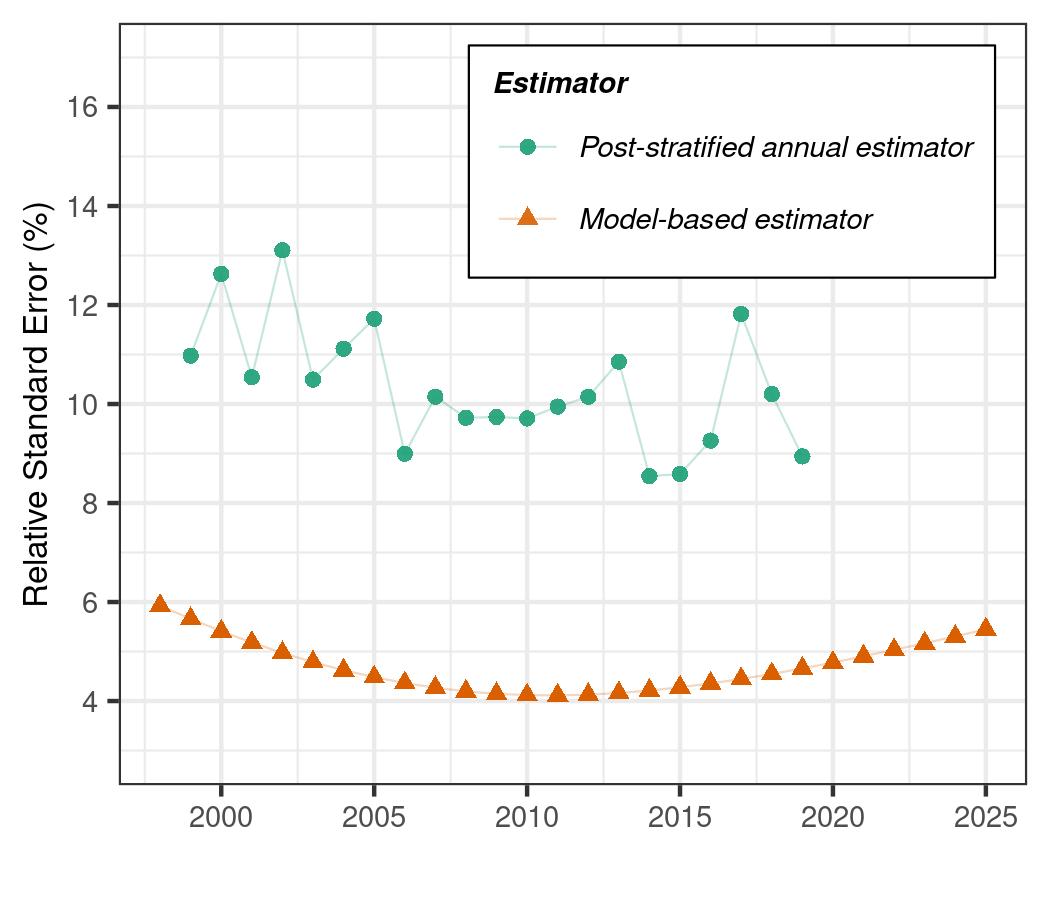}
    \caption{Relative standard error (\%) of model-based and post-stratified estimators of trends total merchantable wood volume on timberland in Washington County, Maine.}
    \label{fig:volume_cv}
\end{figure}

\section*{Results}

Results from design-based and model-based estimators are often not strictly comparable due to fundamental differences in their underlying inferential paradigms, see, e.g., \citet{little2004model}. Of particular importance, design-based estimators can be reasonably assumed unbiased for large samples, whereas model-based estimators cannot be assured to be unbiased \citep{lohr2019sampling}, and in the event of model mis-specification, adverse effects on inference can be substantial \citep{little2004model}. Even among model-based estimators, frequentist and Bayesian inferences yield different interpretation in some cases, see, e.g., \citet{gelmanbda04bda}. Therefore, comparing results derived from these different paradigms, presented in subsequent sections, should be received with an understanding about the respective modes of inference. For example, in some cases we compare design-based estimate derived confidence intervals to Bayesian model-based credible intervals. While it can be convincingly argued such comparisons are not appropriate, we present comparative results to explore general patterns in estimates and highlight estimators' qualities.

\subsection*{County-level forest carbon stocks}
Our results indicate the EBLUP derived from the spatial Fay-Herriot model (described in Eqs \ref{model1a}-\ref{model1c}) offers considerable improvements in precision relative to the post-stratified estimator of county-level forest carbon stocks across much of the CONUS. We present model-based estimates of mean forest carbon density, along with associated estimates of precision, in Figure \ref{fig:smoothed}. Similarly, we map the spatial distribution of the SER in Figure \ref{fig:error}. Finally, we illustrate improvements in relative precision offered by the model-based estimator (i.e., measured by the relative standard error), along a gradient of relative precision in the post-stratified estimator, in Figure \ref{fig:carbon_cv}.

The spatial Fay-Herriot model yields spatially smooth estimates of county-level forest carbon stocks, that generally reflects the distribution of forestland across the CONUS (Figure \ref{fig:smoothed}). The largest estimated forest carbon densities are in the coastal Pacific Northwest, Northern Lake States, and Appalachian regions. In contrast, the smallest estimated forest carbon densities appear in the Southwest, Great Basin, and Northern Plains. We show the relative precision of the model-based estimator generally decreases with estimated mean forest carbon density (Figure \ref{fig:smoothed}; lower precision in counties with low carbon density relative to high carbon density) and with county size (Figure \ref{fig:carbon_cv}; lower precision in small counties relative to large counties). Notably, we show the relative precision of the model-based estimator was generally smallest in the Northern Plains and Southern Lake States regions, likely arising from a combination of small county sizes and relatively low forestland area.

We show the model-based estimator of forest carbon stocks offered the greatest improvements in precision in the coastal Pacific Northwest and eastern US, relative to the post-stratified estimator (Figure \ref{fig:error}). In these regions, the SER commonly fell below 0.5, indicating the standard error of the model-based estimator was less than half that of the post-stratified estimator for a given county. Across the Interior West, in contrast, we show the model-based estimator rarely improved precision by more than 10\% (i.e., SER commonly exceeded 0.9). Further, results presented in Figure \ref{fig:carbon_cv} indicate the model-based estimator generally offered consistent improvements in relative precision over the post-stratified estimator, regardless of the absolute magnitude of the post-stratified estimator's relative precision.  

\subsection*{Trends in merchantable wood volume in Washington County}
Our results indicate the model-based estimator of total merchantable wood volume in Washington County, Maine (approach described in Eqs \ref{model2}-\ref{model2:total}) offers substantial improvements in precision relative to the post-stratified estimator (Figures \ref{fig:volume}-\ref{fig:volume_cv}). Specifically, we show 95\% credible intervals associated with model-based point estimates are consistently narrower than 95\% confidence intervals associated with the post-stratified estimator (Figure \ref{fig:volume}). On average over the period 1999-2019, the relative standard error of the model-based estimator was 55.9\% lower than that of the post-stratified estimator (ranging from 48.9\%-62.4\% lower across all years) (Figure \ref{fig:volume_cv}), indicating the model-based estimator is more than twice as precise as the post-stratified estimator for our domain of interest. Further, consistent alignment of post-stratified and model-based point estimates suggests the model-based estimator is generally unbiased for the domain of interest (Figure \ref{fig:volume}). 

Both approaches indicate that total merchantable wood volume in Washington County has increased considerably over the period 1999-2019 (Figure \ref{fig:volume}). Notably however, the model-based approach yields a smooth, linear trend in total merchantable volume. The post-stratified estimator, in contrast, exhibits large inter-annual variability ($\pm$5-10\% per year, arising from sampling) and pronounced cyclical patterns over the same period (arising from remeasurement of annual panels). Further, the model-based estimator offers an intuitive approach to characterize the magnitude, direction, and statistical significance of temporal trends in our target variable---a feature the post-stratified estimator lacks (absent estimating change from remeasured plots). Specifically, the posterior distribution of the adjusted population-level regression coefficient, $\hat{\beta}$, yields an estimator of average annual change in total merchantable wood volume across our domain of interest. The posterior median of $\hat{\beta}$ was 1,293,900$\mathrm{m}^3 \cdot \mathrm{yr}^{-1}$ (95\% credible interval: 588,000-1,989,000$\mathrm{m}^3 \cdot \mathrm{yr}^{-1}$), indicating a relatively rapid increase in total merchantable wood volume over the last two decades. Further, we show the probability that $\hat{\beta}$ exceeds 0 is $>0.999$, indicating very high certainty in the observed upward trend.  

Finally, the model-based approach offers the ability to forecast changes in our variable of interest, along with associated estimates of uncertainty. We highlight this unique capacity in Figures \ref{fig:volume} and \ref{fig:volume_cv} by predicting total merchantable wood volume, along with estimates of relative precision, over the period 2020-2025---years for which no FIA data has yet been collected/released for our target population. By the year 2025, we estimate, with 95\% probability, that total merchantable wood volume on timberland in Washington County, Maine will range between 124.4-154.7$\mathrm{m}^3$.

\section*{Discussion}
The FIA program operates the largest network of permanent forest inventory plots in the world, making it well suited to provide critical information on US forests over large geographic and temporal domains (e.g., periodic, state-level estimates). However, the program has experienced increased demand for estimates of forest variables for smaller spatial and temporal domains than traditional sample-based estimation approaches can deliver. Providing such estimates without additional investments in field sampling requires adopting alternative estimation approaches. Here, we presented two case studies that demonstrated some aspects of rFIA's potential to simplify application of SAE to data collected by the FIA program, and thus accelerate adoption of such techniques by FIA data users.

First, we estimate contemporary county-level forest carbon stocks across the CONUS using a domain-level spatial Fay-Herriot model (Figure \ref{fig:smoothed}), and show the model-based approach offers considerable gains in precision across the predominately forested regions of the CONUS (Figure \ref{fig:error}). Previous efforts have applied spatial Fay-Herriot models to FIA data to improve precision of estimators of forest density variables \citep{goerndt2011comparison}, private landowner characteristics \citep{ver2017hierarchical}, and forestland removals \citep{coulston2021enhancing}. Domain-level models are particularly useful when inventory plot locations are unknown or measured imperfectly, as spatial auxiliary data need not be associated with plot locations, but rather with domains \citep{rao2015small, mauro2017analysis}. That is, spatial predictors can be used in domain-level models without requiring the release of actual FIA plot locations. We provide all code and data used to develop the domain-level model presented herein in Appendix A, on GitHub (\url{https://github.com/hunter-stanke/FGC_rFIA_SAE}) and at our official website (\url{https://rfia.netlify.app}). Our procedures can be easily adapted for use with alternative target variables, spatial regions, and/or auxiliary data, and we encourage interested users to adapt our code for use in their own applications of domain-level SAE models.

Second, we follow the approach presented in \citet{little2004model} to develop a temporally-explicit unit-level estimator of multi-decadal trends in merchantable wood volume in Washington County, Maine, using a Bayesian multi-level model. We show the model-based approach offered substantial improvements in precision of annual estimates, relative to the traditional, post-stratified approach (Figures \ref{fig:volume}-\ref{fig:volume_cv}). Further, we show the model-based estimator offers an intuitive approach to characterizing the magnitude, direction, and statistical significance of temporal trends, and allows predictions of the target variable to be made for unobserved domains, with associated uncertainty (e.g., forecast change). Unit-level SAE models have been widely applied to FIA data in recent decades \citep{goerndt2011comparison, mcroberts2017multivariate, ohmann2002predictive, babcock2018geostatistical}, and frequently draw from remotely-sensed auxiliary variables to support domain estimation. However, extending the approach presented herein to incorporate spatial auxiliary data will present challenges for most users of FIA data, as neither the true locations of inventory plots, nor the spatial boundaries of strata used for post-stratification are available in the public version of the FIA Database. Nevertheless, the unit-level model presented can be easily adapted for applications involving alternative populations of interest, and might be useful in the detection and characterization of long-term change in forest ecosystems. Further, such models can be used to characterize the status and change in forest variables at spatial and/or temporal domains that are not currently possible using sample-based approaches (e.g., stand-level estimates). 

\subsection*{Role of rFIA in accelerating the adoption of SAE techniques for FIA data}
We posit that rFIA has the potential to simplify the application of model-based SAE techniques to FIA data in three key ways. First, rFIA implements standard, periodic post-stratified estimators---consistent with the estimators implemented by FIA's popular online estimation tool, EVALIDator \citep{evalidator}---within highly flexible, user-defined domains. These direct estimators, along with their associated variances, form the basis for construction of domain-level estimators, as demonstrated by our spatial Fay-Herriot model \citep{fay1979estimates, petrucci2006small, pratesi2008small} of county-level forest carbon stocks. Second, rFIA implements post-stratified estimators for individual annual panels, offering increased temporal specificity over standard periodic estimation approaches (i.e., the temporally-indifferent estimator), and supporting the development of small area estimators that require direct annual estimates of forest variables at aggregate scales. Examples of such temporally-explicit, domain-level estimators include mixed-estimators \citep{van1999modeling} and the spatial-temporal Fay-Herriot model \citep{marhuenda2013small}. Finally, rFIA allows summaries of forest variables to be returned for individual response units (i.e., plot-level) and provides utility functions for extracting design information relevant to particular inventory cycles (e.g., stratum assignments and weights). Together, these data can be used to construct a wide variety of unit-level estimators that acknowledge features of FIA's survey design, as demonstrated in our multi-level model of trends in merchantable wood volume in Washington County, Maine. 

Adoption of SAE methods by FIA data users (particularly new users) is limited more by FIA's complex data structure and survey design than by the availability of tools that implement SAE methods. Thus, we argue the primary benefit of rFIA in accelerating SAE method adoption is its ability to simplify the process of summarizing and formatting FIA data to serve as input to a wide variety of SAE models. There is a large suite of existing, open-source tools that provide generalized implementations of many domain-level and unit-level SAE models. For example, the sae R package \citep{molina2015sae} is specifically designed to implement domain-level SAE models, and we draw from this functionality to develop the domain-level model of forest carbon stocks presented herein. Our intention is not to duplicate efforts of others by implementing common SAE models natively in rFIA, but rather to reduce barriers to the application of such SAE models to FIA data that arise from the complexity of FIA's data structure and sampling design. 

\subsection*{Future extensions of rFIA}
Current efforts to extend rFIA include the implementation of a suite of model-based time-series estimators that aim to improve the precision of annual estimates of forest variables, thereby increasing the relevance of FIA data for change detection, characterization, and attribution. Specifically, we aim to provide an intuitive implementation of Van Deusen's mixed-estimator \citep{van1999modeling}, which was recently shown by \citet{hou2021updating} to offer considerable improvements in the precision of annual FIA-based forest land area estimates, at both the state- and county-levels. Further, we aim to provide an alternative Bayesian estimator of annual trends in forest variables based on a measurement error model (e.g., similar to Bayesian meta-analysis \citep{sutton2001bayesian}). Notably, both approaches effectively smooth annual, post-stratified estimates of forest variables, and hence are compatible with FIA's existing survey design and database structure.

\subsection*{Acknowledgements}
This work was supported by: National Science Foundation grants DMS-1916395, EF-1253225, EF-1241874; USDA Forest Service, Region 9, Forest Health Protection, Northern Research Station; US National Park Service; and Michigan State University AgBioResearch.

\subsection*{Author Contributions}
H.S. and A.O.F designed research; H.S. performed research; H.S. analyzed data; H.S., A.O.F, and G.M.D wrote and reviewed the manuscript.

\clearpage

\printbibliography

\end{document}